\documentclass[modern,astrosymb]{aastex631}
\usepackage{xspace}

\newcommand{\code}[1]{{\texttt{#1}}}

\newcommand{\by}{\mathbf{y}}
\newcommand{\bx}{\mathbf{x}}
\newcommand{\bs}{\mathbf{s}}
\newcommand{\bw}{\mathbf{w}}
\newcommand{\bh}{\mathbf{h}}
\newcommand{\bk}{\mathbf{k}}
\newcommand{\ba}{\mathbf{a}}
\newcommand{\oii}{[O\,{\footnotesize II}]}
\newcommand{\sii}{[S\,{\footnotesize II}]}
\newcommand{\nii}{[N\,{\footnotesize II}]}
\newcommand{\spender}{\textsc{spender}\xspace}
\newcommand{\btheta}{{\boldsymbol{\theta}}}

\usepackage{amsmath}
\usepackage{bm}
\usepackage{booktabs}

\shorttitle{Autoencoding Galaxy Spectra}
\shortauthors{Melchior et al.}

\graphicspath{{./}{figures/}}

\begin{document}

\title{Autoencoding Galaxy Spectra I: Architecture}

\correspondingauthor{Peter Melchior}
\email{peter.melchior@princeton.edu}

\author[0000-0001-9592-4190]{Peter Melchior}
\affiliation{Department of Astrophysical Sciences, Princeton University, Princeton, NJ 08544, USA}
\affiliation{Center for Statistics \& Machine Learning, Princeton University, Princeton, NJ 08544, USA}

\author[0000-0002-1001-1235]{Yan Liang}
\affiliation{Department of Astrophysical Sciences, Princeton University, Princeton, NJ 08544, USA}

\author[0000-0003-1197-0902]{ChangHoon Hahn}
\affiliation{Department of Astrophysical Sciences, Princeton University, Princeton, NJ 08544, USA}

\author[0000-0003-4700-663X]{Andy Goulding}
\affiliation{Department of Astrophysical Sciences, Princeton University, Princeton, NJ 08544, USA}

\begin{abstract}
  We introduce the neural network architecture \spender as a core differentiable building block for analyzing, representing, and creating galaxy spectra.
  It combines a convolutional encoder, which pays attention to up to 256 spectral features and compresses them into a low-dimensional latent space, with a decoder that generates a restframe representation, whose spectral range and resolution exceeds that of the observing instrument. The decoder is followed by explicit redshift, resampling, and convolution transformations to match the observations.
  The architecture takes galaxy spectra at arbitrary redshifts and is robust to glitches like residuals of the skyline subtraction, so that spectra from a large survey can be ingested directly without additional preprocessing.
  We demonstrate the performance of \spender by training on the entire spectroscopic galaxy sample of SDSS-II; show its ability to create highly accurate reconstructions with substantially reduced noise; perform deconvolution and oversampling for a super-resolution model that resolves the \oii\ doublet; introduce a novel method to interpret attention weights as proxies for important spectral features; and infer the main degrees of freedom represented in the latent space. We conclude with a discussion of future improvements and applications.
\end{abstract}

\keywords{galaxies: statistics -- techniques: spectroscopic}

\section{Introduction}

Spectroscopy is critical to understand the physical processes that happen in galaxies across cosmic time.
But despite the availability of millions of galaxy spectra from large surveys and dedicated programs, we still lack models that capture their full distinctiveness and diversity, especially when redshift evolution is to be taken into account.
Theoretical models cannot reproduce high-quality observed spectra, not even when restricted to specific galaxy subpopulations \citep[e.g.][]{tojeiro2011}. 
Many physics-based approaches also treat separately the continuum from stellar emission 
and emission lines from nebular emission \citep{Baldwin1981-ww, Kewley2019-pu}.
This practice creates a disconnect between a galaxy's stellar population and its gas content, which then affects the capabilities of subsequent analysis efforts \citep{cappellari2017, leja2017}.

On the other hand, data-driven spectrum models have been limited to high-quality observations of nearby galaxies, where the cosmological redshift can be ignored \citep{Moustakas2006-uy, Brown2014-im}, or to a reduced wavelength range that is accessible for all galaxies after they have been transformed back to restframe \citep{Yip2004-ob, Portillo2020-yr,Teimoorinia2022}.
Either way reduces the number of usable galaxies or the wavelength range over which galaxies can be useful.
The assumption that any given spectrum can be represented by a linear combination of a small number of basis vectors \citep{Yip2004-ob} or prototypical templates \citep{Calzetti1994-sr, Kinney1996-da} further limits the complexity of data-driven spectrum models.
Despite their simplicity, linear models are widely used for inferring redshifts from galaxy spectra \citep{bolton2012, ross2020} or broadband photometry \citep{Benitez2000-dr, Brammer2008-fj}, as well as for the generation of mock spectra from large cosmological simulations \citep{fagioli2018, wechsler2021}.

In summary, neither data-driven nor theoretical models currently capture the full information content of galaxy spectra.
This means, e.g., that we cannot robustly marginalize over galaxy properties when inferring spectroscopic redshifts, causing catastrophic outliers from line misidentification \citep{Cunha2014}. It is similarly difficult to reliably assess whether a spectrum as a whole or over some smaller wavelength range shows unusual behavior \citep{Lochner2021}, or estimate the effective number of clusters or degrees of freedom in galaxy spectra \citep{Rahmani2018, Fraix-Burnet2021}.

However, the widespread adoption of template libraries suggests that galaxy spectra in fact occupy a low-dimensional manifold.
More explicitly, \cite{Portillo2020-yr} demonstrated that a high-fidelity reconstruction can be achieved by an autoencoder architecture \citep[AE,][]{Hinton1994-gu, Kingma2013-ai} with a latent space of just six dimensions.
\citet{Teimoorinia2022} improved upon that work by introducing convolutional elements into the AE to aid the extraction of correlated features from the spectra.
We further advance this approach with a specifically designed architecture that combines an attentive convolutional encoder with an explicit redshift transformation after the decoder.
It allows the exploitation of the full spectrum of all galaxies in a survey to form a super-model that exceeds the wavelength range and spectral resolution of any individual spectrum.

\section{Data}
\label{sec:data}

We retrieve $\approx 650,000$ spectra from the Main Galaxy Sample of the Sloan Digital Sky Survey \citep[SDSS-II,][]{strauss2001}, which is magnitude limited at Petrosian $r < 17.77$ and  covers redshifts up to $z_\mathrm{max}\approx 0.5$, with the large majority at $z\lesssim 0.25$.
We use calibrated spectra, inverse variance weights, and masks from the spectroscopic reduction of SDSS Data Release 16 \citep{ahumada2020}.
We select spectra classified as galaxies, with valid redshift estimates and redshift errors $\sigma_z < 10^{-4}$, and without quality flags that would indicate
issues with spectrophotometric calibration or sky subtraction. 
After padding into a fixed length array, each spectrum has $M=3921$ spectral elements and covers the wavelength range $\lambda = 3784 \dots 9333$\,\AA.
We normalize the spectra by dividing out the median spectral flux over the range $\lambda=5300\dots5850\,\mathrm{\AA}/(1+z)$, choosing a restframe normalization to avoid a redshift dependence of the signal amplitude (see \autoref{sec:latents} for more discussion) and this specific region because it is devoid of strong emission and absorption lines.
We also add to the masks regions of 5\,\AA\ around the 100 strongest sky lines; doing so should account for any residuals of the skyline subtraction procedure in the DR16 pipeline.
In masked areas, we set the weights to zero, but leave the spectra unchanged.

\section{Architecture}
\label{sec:architecture}

\begin{figure*}[t]
    \includegraphics[width=\textwidth]{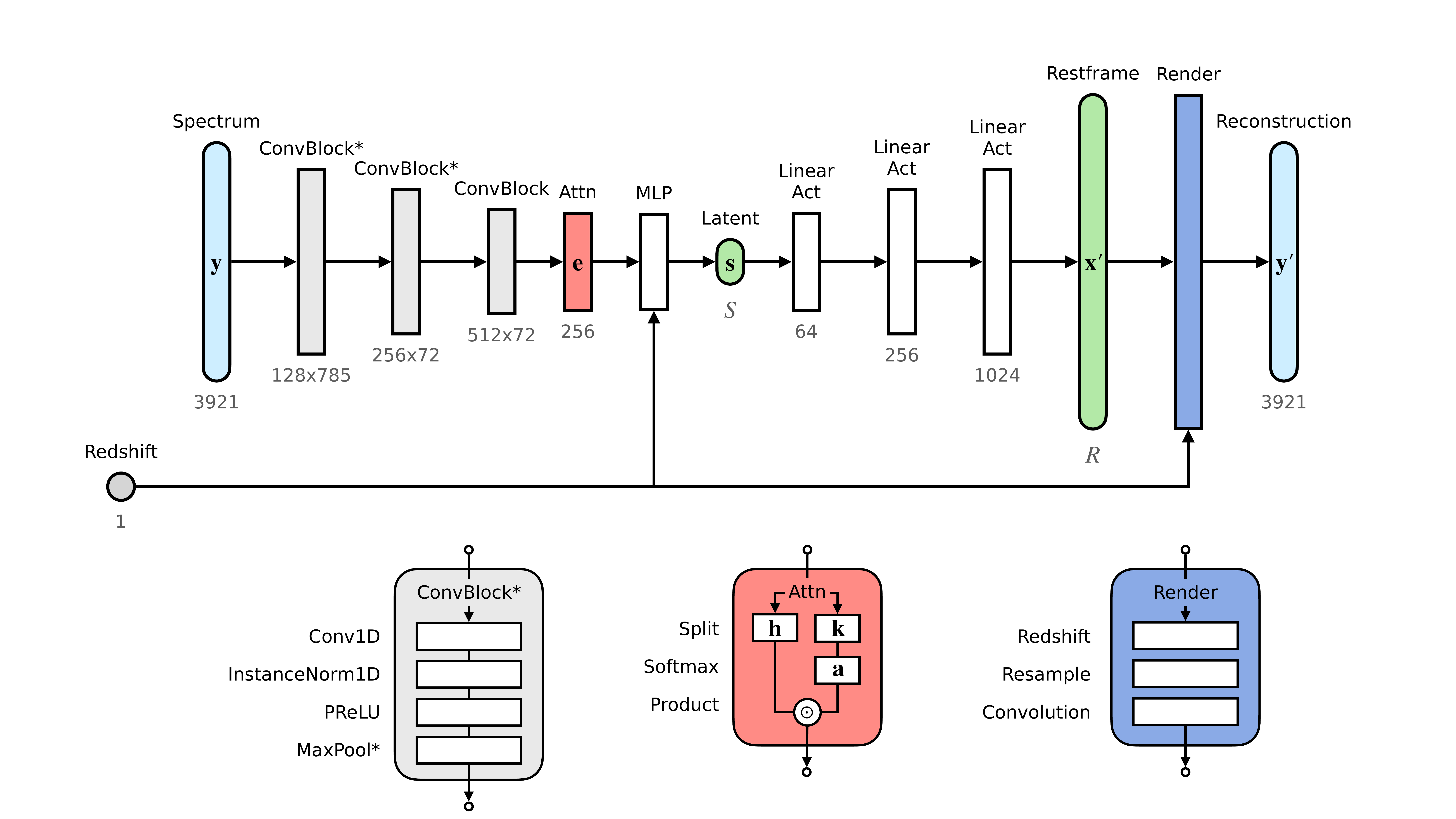}
    \caption{The autoencoder architecture of \spender with an attentive convolutional encoder and an explicit redshifting, resampling, and convolution transformation after the decoder.}
    \label{fig:architecture}
\end{figure*}

\autoref{fig:architecture} shows a sketch of the \spender architecture, whose elements we discuss in detail below.
As usual for an AE model, it is trained end-to-end with a (weighted) MSE loss
\begin{equation}
    \label{eq:loss}
    l(\btheta) = \frac{1}{2 L}\sum_m^M \left(\bw \odot (\by - \mathbf{f}_{\btheta}(\by))^2\right)_m,
\end{equation}
where $\by\in\mathbb{R}^{M}$ denotes an observed spectrum, $\by^\prime\equiv\mathbf{f}_\btheta(\by)$ its reconstruction by the autoencoder with parameters $\btheta$, $w_m\equiv1/\sigma_m^2$ the inverse variance weight of the $m$-th spectral element, and $\odot$ the element-wise multiplication.
This loss is proportional to the log-likelihood for normal-distributed data and explicitly allows us to incorporate heteroscedastic uncertainties and variable masking through element-wise weights.

\subsection{The Encoder}
Our central design choice is to leave the data in the given form, i.e. in the observed frame, not in restframe.
Doing so preserves all aspects---and the entire wavelength range---of the observed spectra.
But this decision poses a greater challenge for the encoder because it is now confronted with varying locations for spectral features.
A common machine learning technique for searching patterns across the data domain is called \emph{attention}, which involves learning to weigh features from the input data that are beneficial for subsequent tasks (see e.g. \citet{Chaudhari2021-jr} for a recent review of attention methods).
As we expect such features to be correlated between sharp lines or breaks and the continuum, we chose the convolutional encoder architecture from \cite{Serra2018-lw} to extract latent parameters from observed spectra.
The architecture starts with three convolutional layers with progressively wider kernel sizes (5, 11, 21), trainable PReLU activations \citep{He2015}, and max-pooling, which translates $M=3921$ spectral elements into 512 channels for 72 wavelength segments.%
\footnote{The stride of the MaxPool layers was not specified by \citet{Serra2018-lw}. We adopted the conventional approach to match the preceding convolution kernel sizes, which results in a receptive window of 1208 spectral elements \citep{Araujo2019} and a relatively strong wavelength compression. This choice can be tuned if larger receptive windows or less spectral compression is desired.}
It then applies attention in wavelength direction to these channels, i.e. it splits the channels into two parts, $\mathbf{h}$ and $\mathbf{k}$ ($\in\mathbb{R}^{256 \times 72}$), and combines them as
\begin{equation}
\label{eq:attn}
\mathbf{e}=\bh\cdot\mathrm{softmax}(\bk)\equiv \bh\cdot\ba,    
\end{equation}
where the dot product and the softmax operate on the last, i.e. the wavelength dimension. 
The vector $\ba$ contains the \emph{attention weights}, indicating whether and where relevant signals have been found, so that their corresponding values are promoted to the attended features $\mathbf{e}$.
This architecture is capable of accounting for the apparent shift of spectral features in galaxies at different redshifts.
It behaves similar to traditional redshift estimation techniques that scan for particular spectral lines  \citep[e.g.][]{SubbaRao2002-gb} and, because of the wide convolution kernels, naturally folds in continuum features to form a highly informative latent representation.

\subsection{The Decoder}
\label{sec:decoder}

The decoder is a simple expansive 3-layer Multi-Layer Perceptron (MLP) with (64, 256, 1024) hidden dimensions, which generates an internal \emph{restframe} representation $\bx\in\mathbb{R}^R$ of the spectrum, with $R\geq M$. This internal model is then analytically redshifted to the known galaxy redshift $z$ and resampled to the observed spectrum wavelengths and resolution $M$.
A convolution with the line spread function (LSF) can also be performed at this stage (we will discuss this option further in \autoref{sec:superres}). 
Applying the redshifting transformation explicitly in the generator part of the autoencoder removes the ambiguity between the location of spectral features in restframe $\bx$ and in the observed frame $\by$.
It is thus much easier to train the critical aspect, namely the spectral encoding of \emph{restframe} features, without also having to learn the analytically known effects of redshift.
A similar approach was employed by \cite{Lanusse2021-gn} for encoding galaxy images under variations of the optical point spread function.

We choose the activation function, proposed by \citet{Alsing2020-ps} specifically for galaxy spectra,
\begin{equation}
\label{eq:speculator}
   a(\bx) = \left[\boldsymbol{\gamma} + (1-\boldsymbol{\gamma})/(1+\exp(-\boldsymbol{\beta}\odot\bx))\right]\odot\bx.
\end{equation}
Compared to a more conventional ReLU, this function includes additional trainable parameters, which help generate spectra with flat and sharp features.
By initializing $\boldsymbol{\beta}=\mathbf{1}$ and $\boldsymbol{\gamma}=\mathbf{0}$, the MLP produces values very close to $\mathbf{0}$. We add $\mathbf{1}$ to the output of the last activation because our observed spectra have been normalized to a unit median.
The decoder thus has to learn to deviation from a flat median spectrum.

To generate a complete redshifted spectrum, the restframe model $\bx$ must encompass the maximum range of restframe wavelengths covered by any spectrum in the dataset.
Considering SDSS galaxy spectra up to $z_\mathrm{max}\approx 0.5$ forces us to lower $\lambda_\mathrm{min}$ to $3784\,\mathrm{\AA}/(1+z_\mathrm{max})$.
By the same token, a high-fidelity resampling operation suggests that the restframe has at least $M(1+z_\mathrm{max})$ linearly spaced spectral elements.
We perform the resampling step as a linear interpolation, which appears sufficient for the SDSS native resolution and LSF.
Should a higher-fidelity resampling be desired, one can combine it with the LSF convolution using a one-dimensional version of the method by \citet{Joseph2021}.
If the architecture can successfully be trained, the resulting restframe model will have super-resolution over the extended wavelength range. 
We will exploit this property in \autoref{sec:superres}.

\subsection{Additional Features}

We include an extra MLP on the encoder side, which further compresses the attended features $\mathbf{e}$ \emph{and the redshift} to latent variables $\bs\in\mathbb{R}^S$.
Providing the redshift as an input variable allows the encoder to learn the relation of spectral features---their strength as well as their overall presence or absence---with redshift.
Without that, it would be impossible to learn a spectrum encoding that generalizes across redshifts if some important spectral feature can get redshifted out of the observable wavelength window, like H$\alpha$ at $z\approx 0.4$.

The \spender architecture provides a clear assignment of responsibilities across the AE modules.
Reading \autoref{fig:architecture} from right to left, the decoder needs to generate a restframe spectrum from a low-dimensional latent representation.
The encoder is thus forced to find such a restframe representation when given an observed, redshifted spectrum and the redshift.

Following a suggestion in \citet{Portillo2020-yr}, we experimented with including data weights $\bw$ in the encoding process.
One option is to compress the weight vector with a CNN encoder and use the compressed representation as additional input variables for the encoder MLP.
Doing so would not encode the location of high and low weights, only their relative strengths and positions.
Instead, we computed attention weights $\mathbf{k}_\bw$ of $\bw$ in the same way, but by a different CNN encoder, as for the spectra. Multiplying the weights, i.e. computing the attended features as $\mathbf{e}_\bw=\mathbf{h}\cdot\mathrm{softmax}(\mathbf{k} \odot \mathbf{k}_\bw)$,
has the effect of modulating attention, ideally reducing it for spectral features with low weights.
However, we found that the resulting models achieved essentially the same final loss $L$, but the attended features where much less directly tied to prominent spectral lines and breaks (see \autoref{sec:attention}).
Instead, a large amount of attention has been expended on regions with strong sky lines.
Moreover, even without encoding weights, the reconstructions $\mathbf{f}_\btheta(\by)$ are robust around areas with large artifacts, as we will discuss in \autoref{sec:inspection}.
We see no benefit of providing weights to the encoder and therefore proceed without doing so.

\section{Results}
\label{sec:results}

\begin{table}[t]
    \centering
    \begin{tabular}{lcccc}
    \toprule
    S & \multicolumn{2}{c}{LR: $R=5881$} & \multicolumn{2}{c}{SR: $R=11762$, LSF5}\\
    $S$ & Training Loss & Validation Loss & Training Loss & Validation Loss\\
    \cmidrule(lr){2-3}\cmidrule(lr){4-5}
    2 & 0.426 & 0.430 & 0.425 & 0.430\\
    4 & 0.394 & 0.396 & 0.392 & 0.395\\
    6 & 0.385 & 0.388 & 0.385 & 0.387\\
    8 & 0.382 & 0.383 & 0.381 & 0.383\\
    10 & 0.380 & 0.381 & 0.379 & 0.381\\
    \bottomrule
    \end{tabular}
    \caption{Average MSE loss values for an ensemble of 5 \spender models after 100 training epochs as a function of the dimensionality of the latent space $S$. Loss values below 1 indicate that the average per-element error is smaller than the noise level. The first set of models use the lowest resolution (LR) with $R=5881$ necessary to prevent undersampling the SDSS spectra when redshifted up to $z_\mathrm{max}=0.5$ (cf. \autoref{sec:decoder}). The second set of models use super-resolution (SR) by factor 2 and simultaneously fit for an unknown line spread function kernel of width 5 (see \autoref{sec:superres}). 
    \label{tab:loss}}
\end{table}

We implement the \spender architecture introduced in \autoref{sec:architecture} with \code{pytorch} \citep{pytorch}, and train it with 70\% of the parent sample for 100 epochs with the Adam optimizer \citep{Kingma2015-pq}, a learning rate of $10^{-3}$, and the 1Cycle schedule \citep{Smith2017} on a NVIDIA V100 GPU.
Training takes approximately one hour.
The remaining 30\% of the data are split evenly in validation and test samples.
Column 3 of \autoref{tab:loss} shows the weigthed MSE loss for the training and validation samples as a function of the dimensionality of the latent space for the model with $R=5881$---the minimum resolution high enough to prevent undersampling of the observed model caused by redshifting.
We confirm the findings of \citet{Portillo2020-yr} that models of galaxy optical spectra with fidelity much better than the noise level (MSE $< 1$) can be achieved with few latent parameters, with $S=6$ apparently providing a good trade-off between reconstruction fidelity and parsimony.
We also see that training loss is only mildly smaller than validation loss, suggesting that overfitting does not pose a noticeable problem.
Changing the learning rate has little effect, and the spread between different models in terms of the final loss is on the order of $10^{-3}$.
Experimenting with a leaky ReLU activation in the decoder yields only marginally inferior results to the activation function in \autoref{eq:speculator}.
We conclude from this that the training procedure is overall stable, and the final loss likely dominated by the size, diversity, and effective noise level of the training data.

\subsection{Qualitative Inspection}
\label{sec:inspection}

\begin{figure*}[t]
    \includegraphics[width=\textwidth]{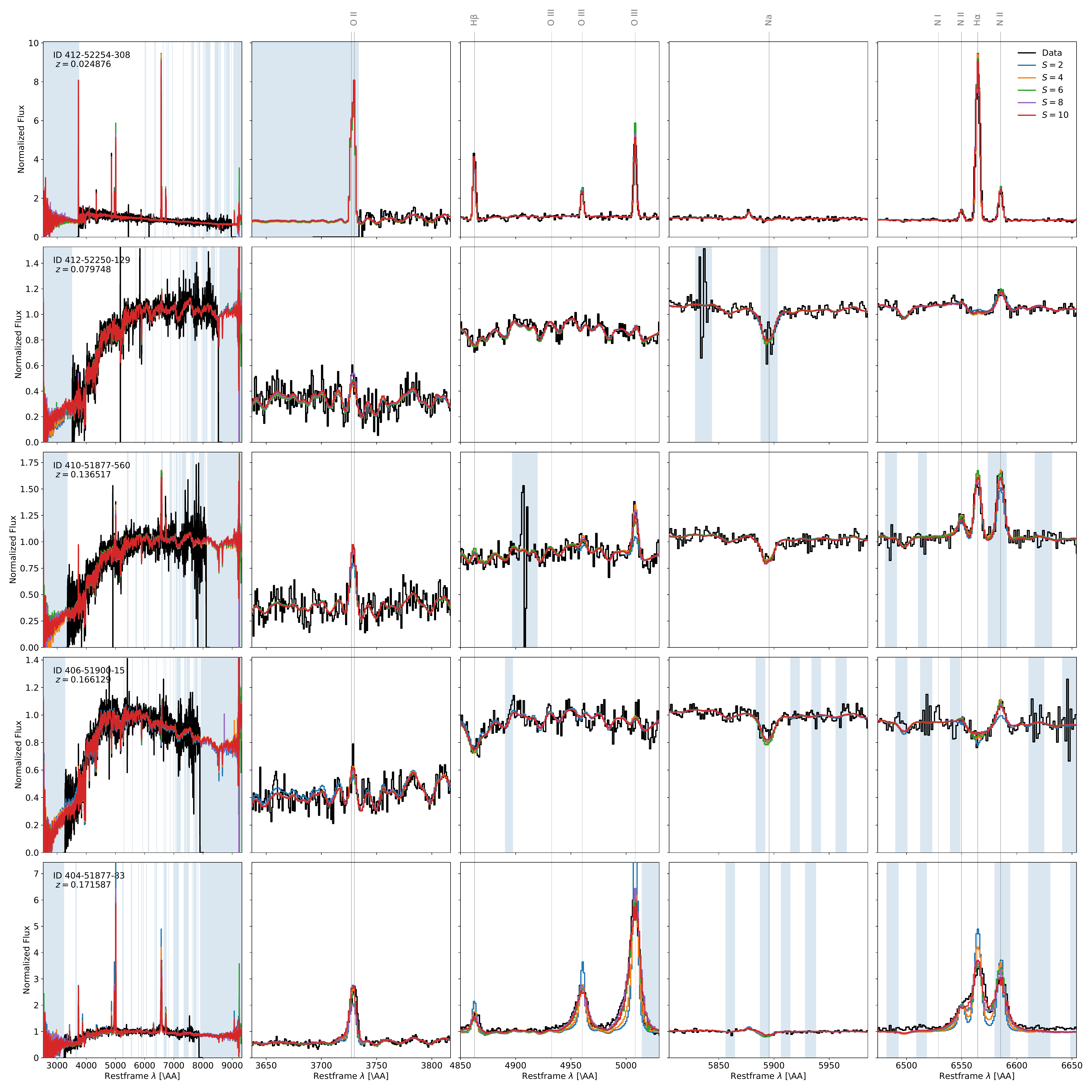}
    \caption{Example restframe spectra, ordered by redshift (\emph{top to bottom}), and their reconstruction by 5 \spender LR models with different latent space dimensionality $S$. The left column shows the entire spectrum. Other columns zoom in to specific emission and absorption lines. Wavelengths that are unobserved or masked by the processing pipeline are shown in light blue shading.}
    \label{fig:examples}
\end{figure*}

\autoref{fig:examples} shows the original SDSS and reconstructed spectra from \spender in the lower-resolution setting with $S\in\{2,4,6,8,10\}$ in the latent variables for several galaxies from the withheld test sample, spanning $z=0\dots0.2$ that is best covered by the SDSS MGS.
All \spender models provide an excellent fit to the data, reducing the effective noise levels by about a factor of 2.
They are essentially indistinguishable from each other, except in the region of strong emission lines, where larger latent space dimensionality improves modeling fidelity.
This is most evident from the recovery of the highly variable behavior of the H$\alpha$-\nii\, complex (last column) for a starforming (first row) to a quiescent galaxy (third row), even in the presence of substantial line broadening (last row).

As is visible from the left column of \autoref{fig:examples}, the \spender restframe model exceeds the wavelength range of any observed spectrum, inferring missing parts from similar galaxies at different redshifts.
The successful establishment of an extended restframe model is most obviously confirmed by the prediction of the expected but unobserved \oii$\lambda\lambda3726,3729$ emission in the first example spectrum of \autoref{fig:examples}.

The \spender models yield good reconstructions despite glitches near and directly at important spectral features (e.g. the Na\,{\footnotesize I} absorption in second example spectrum), demonstrating its robustness to observational artifacts.
Furthermore, the model is robust to higher noise levels (as in the third example).
Both findings are remarkable because neither weights $\mathbf{w}$ nor masks are passed to the encoder.
This robustness suggest that the encoder utilizes many correlated features, e.g. the overall shape of the continuum.
During training, when the statistical weights are available (cf. \autoref{eq:loss}), the encoder evidently learns how to recognize and combine them suitably for good reconstruction fidelity.

\subsection{Super-resolution}
\label{sec:superres}

We can push the excellent modeling capabilities of \spender even further.
As the decoder has an explicit resampling operation (cf. \autoref{sec:decoder}), we are free to choose both the properties of the restframe and the observed frame at will. We could mimic how a SDSS spectrum would appear to a different instrument. Alternatively, we can also chose to increase effective resolution of the restframe model.
Given that the models with $R=5881$ effectively suppress the noise by about a factor of 2, it is reasonable to expect that we can increase the resolution by a similar factor.

But at this point we need to acknowledge the presence of the LSF that we have ignored so far.
The LSF broadens the observed spectra already at native instrument resolution---the effective width of the LSF kernel exceeds one pixel of the spectrograph---which means increasing the resolution now would be pointless: we would not actually resolve finer features, just get longer correlations among smaller pixels. 
We thus perform an additional convolution with the LSF after the resampling step, which forces the decoder to learn the LSF-deconvolved restframe representation.
However, we could not find any publicly available information about the shape of the LSF kernel for SDSS-II spectra.
We therefore set up a LSF kernel of size 5, and learn the average shape of the kernel directly from the data, together with all other parameters of the model.
Doing so sounds impossible because the inverse problem is degenerate:
any observed width of a spectral line could be attributed to its intrinsic width or the width of the LSF kernel.
But the exact degeneracy is broken in our data set as it comprises spectra at different redshifts.
Redshifting causes an apparent stretch of the intrinsic line width but leaves the LSF unchanged.

Training a set of new models with $R=11762$ in exactly the same way as the previous ones produces the MSE losses listed in the right columns of \autoref{tab:loss}.
The fidelity is very slightly improved compared to the lower-resolution models, a mild indication that the generative model benefits either from acknowledging the presence of the LSF or from reducing resampling errors by a higher resolution, or both.
Training times for this model are approximately two hours.

\begin{figure*}[t]
    \includegraphics[width=0.33\textwidth]{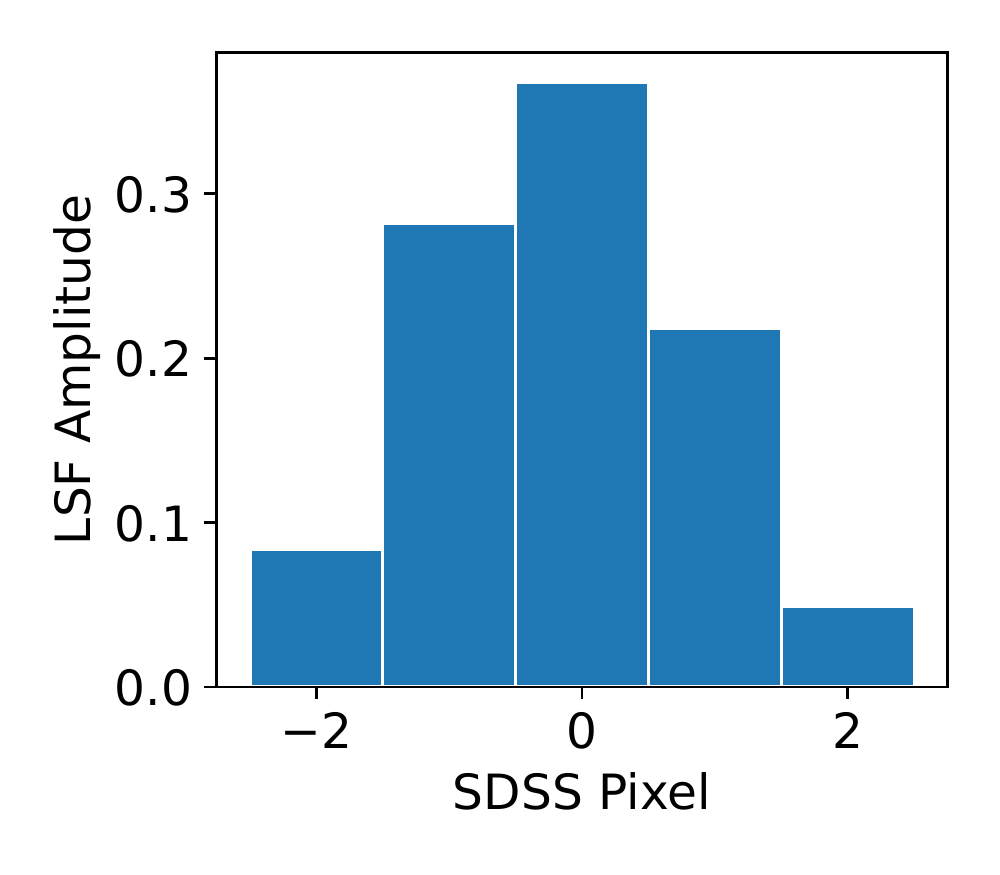}
    \includegraphics[width=0.66\textwidth]{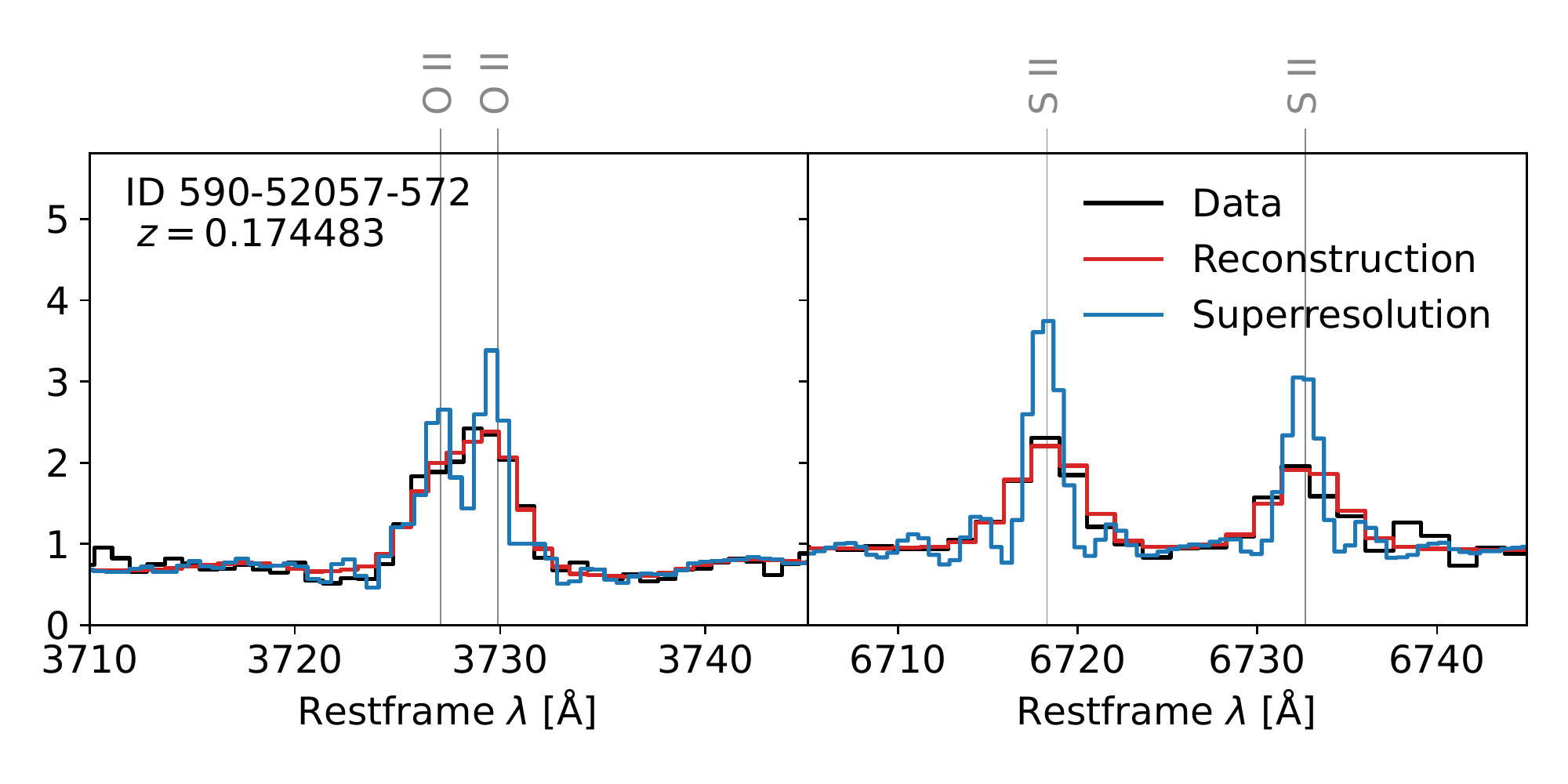}
    \caption{\emph{Left:} 5 pixel wide SDSS-II LSF kernel learned from the data. \emph{Right:} SDSS spectrum in the regions of the \oii$\lambda\lambda3726,3729$ and \sii$\lambda\lambda6716,6731$ doublets. The \spender SR model resolves the \oii\ doublet that is heavily blended in the observation by increasing the restframe resolution by a factor of 2 and deconvolving from the LSF model.}
    \label{fig:superres}
\end{figure*}

The left panel of \autoref{fig:superres} shows the learned LSF shape. Aggregated of the entire data set, it represents the average amount of broadening that does not depend on redshift or specific spectral features. 
As expected, the kernel is peaked in the center and largely symmetric.
These properties are not enforced during training, exhibiting them must therefore be a consequence of the broken degeneracy in the data.

The super-resolution models are overall noisier than the lower resolution ones we inspected in \autoref{sec:inspection}.
This is the expected result of increasing the resolution while attempting to deconvolve noisy data.
Super-resolution of factor 4 proved unadvised with this data set.
However, in regions of prominent features super-resolution is effective.
The right panel of \autoref{fig:superres} focuses on the \oii$\lambda\lambda3726,3729$ doublet, whose lines are heavily blended in the observed SDSS data and are clearly separated in the \spender super-resolution model.

We measure the equivalent widths for both of the lines in the doublet with a Gaussian peak and sloped continuum fit, and find \oii$\lambda$3729/\oii$\lambda$3726 $= 1.27 \pm 0.13$, indicative of relatively low electron density \citep{Osterbrock1974}.
To further determine whether the doublet properties are plausible, we exploit that the \oii\ doublet line ratios are strongly correlated with those of the \sii$\lambda\lambda6716,6731$ doublet \citep{Zeippen1982, Wang2004}, which is accessible for this galaxy (shown in the right panel of \autoref{fig:superres}).
The corresponding ratio \sii$\lambda6716$/\sii$\lambda6731 = 1.35\pm 0.09$, measured directly from the SDSS spectrum, is indeed entirely consistent with the predicted \oii\ doublet line ratios from the super-resolved restframe model.

It is important to realize that the recovery of the doublet is only possible because \spender learns from many similar galaxies, for which the exact positions of the lines with respect to the wavelength bins varies with their redshifts.
The collective loss is marginally lower when representing this region of the spectrum with a doublet instead of a single peak.
The flip side of this argument suggest caution when measuring spectral features from the super-resolution model as they may be prominently biased towards the most common realizations of such features in the training data.

\subsection{Attention}
\label{sec:attention}

The mechanism by which important features are recognized is the attention module in \autoref{fig:architecture}.
It splits the compressed CNN channels into a set of values $\bh$ and keys $\bk$, and turns the latter into probabilistic weights $\ba$, indicating what is important for reconstructing the input spectrum.
Every spectrum creates $\ba\in\mathbb{R}^{256\times72}$, i.e. an attention weight for every channel and every wavelength segment.
The final dot product in \autoref{eq:attn} applies these weights over all compressed wavelength segments.
It can be thought of as a conventional search for spectral features like the 4000\,\AA\ break or the H$\alpha$ line, with the difference that what is considered important is learned in an unsupervised fashion.

\begin{figure*}[t]
    \centering
    \includegraphics[width=0.48\linewidth]{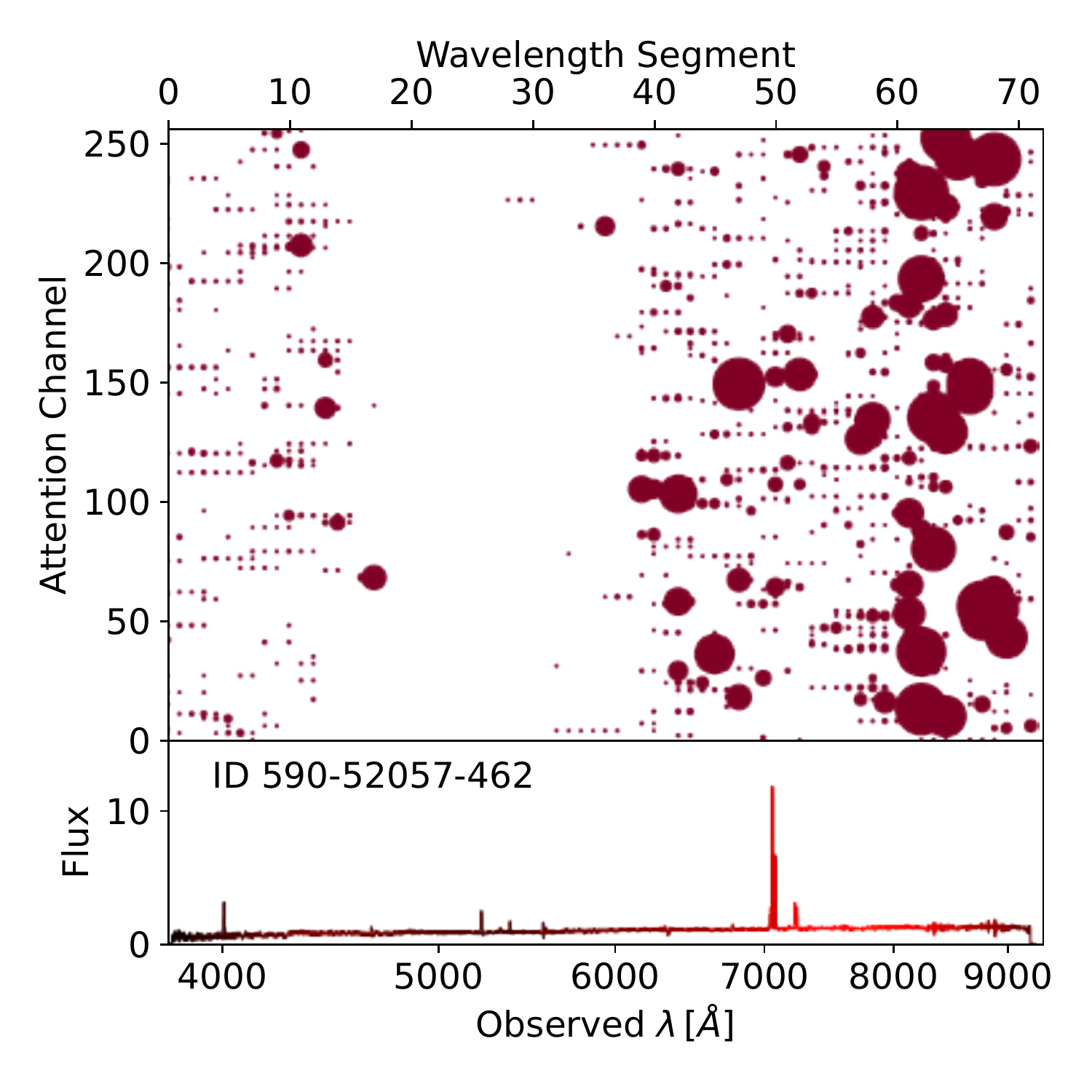}
    \includegraphics[width=0.48\linewidth]{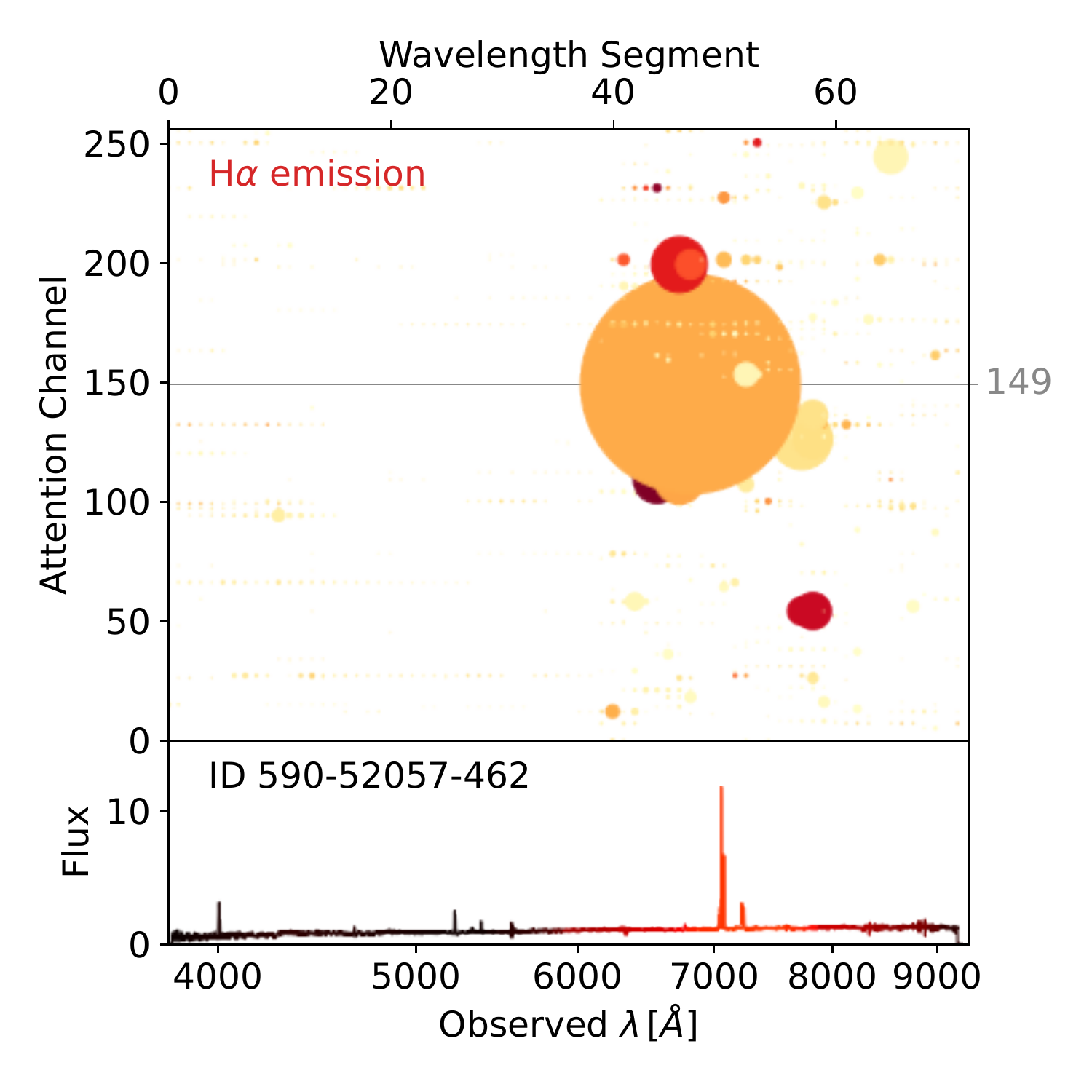}\\
    \includegraphics[width=0.48\linewidth]{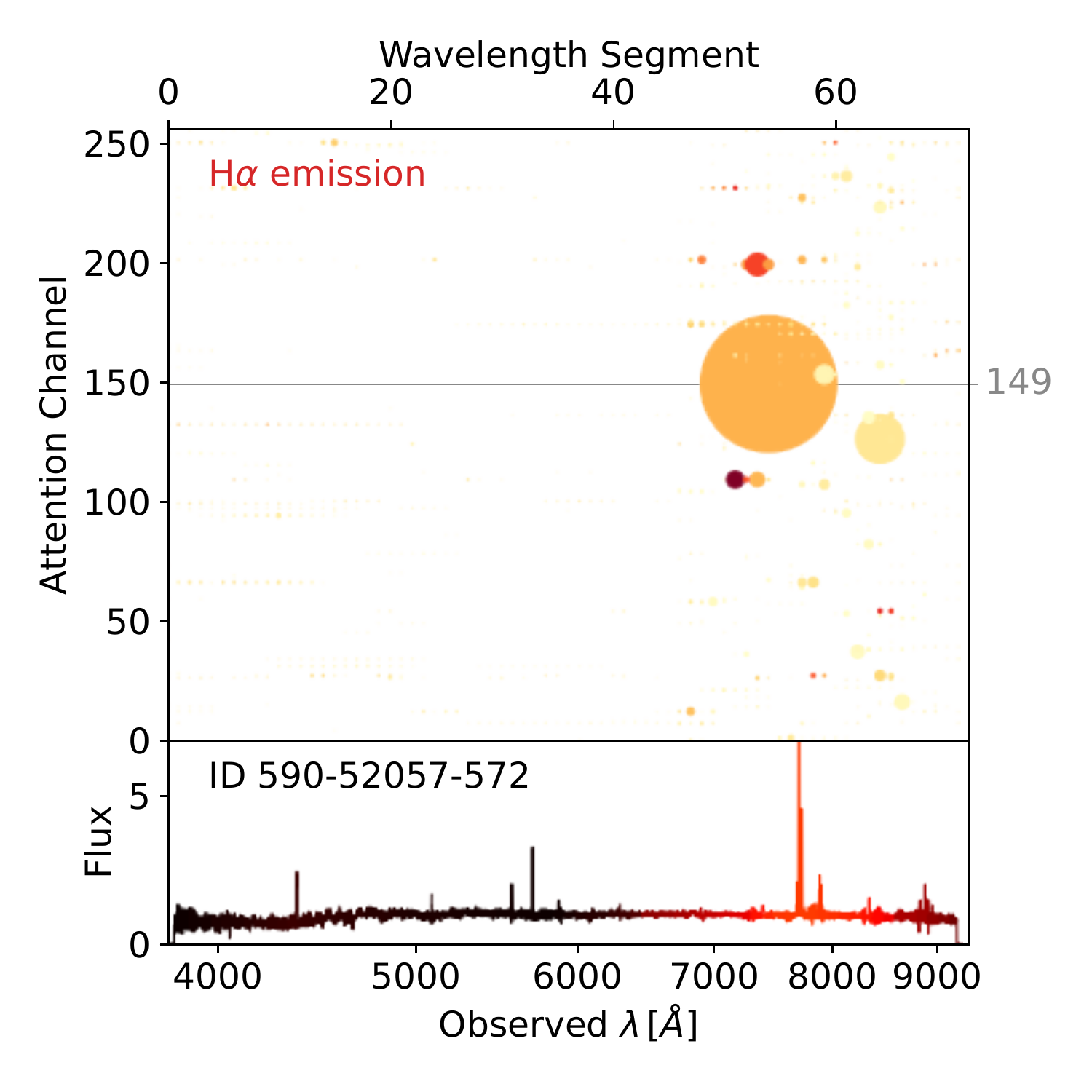}
    \includegraphics[width=0.48\linewidth]{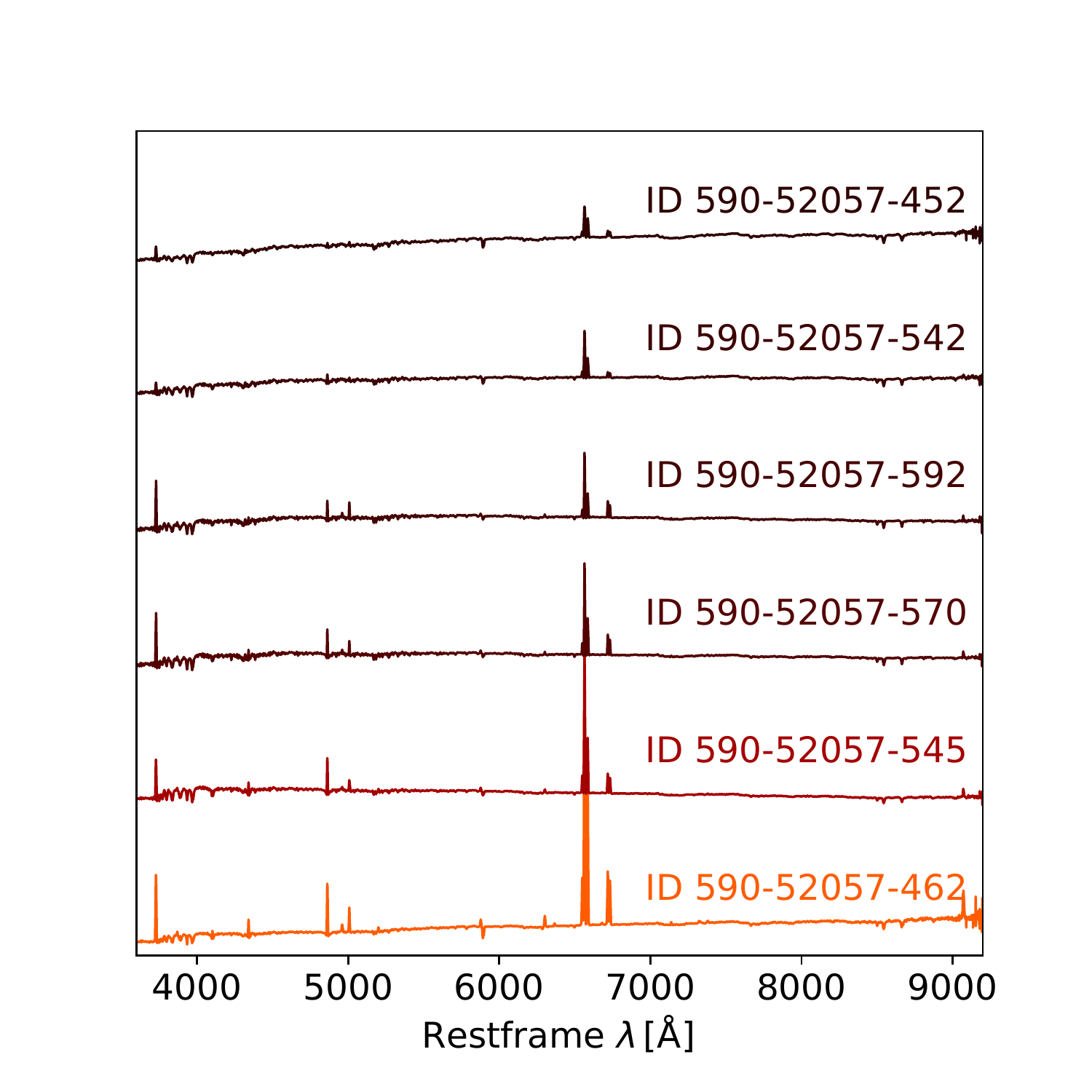}
    \caption{\emph{Top left:} Direct attention weights $\ba$ of the \spender LR-10 model for the spectrum shown in the lower subpanel. Marker sizes are proportional to weight activations. \emph{Top right:} Grad-FAM attention weights $\mathbf{ga}$ responsible for the peak height of the H$\alpha$ line (cf. \autoref{eq:grad-fam}). Colors indicate gradient amplitudes, marker sizes increased for clarity. Attention channel 149 is most highly activated, especially in the wavelength segment that contains the H$\alpha$ line. \emph{Bottom left:} Same as top right, but for a higher redshift galaxy. Channel 149 remains most highly activated, finding H$\alpha$ in a larger wavelength segment. \emph{Bottom right:} Six galaxies from the same SDSS plate, ordered by Grad-FAM activation of channel 149, which serves as a proxy for H$\alpha$ line strength.}
    \label{fig:attn}
\end{figure*}

We expected that significant spectral features have their own attention channels.
Identifying them is not trivial because there is a multitude of aspects in any spectrum that are being attended to in order to create a high-fidelity reconstruction.
The attention weights for a single example galaxy are shown in the top left panel of \autoref{fig:attn}.
We see a large number of activations, with surprisingly many channels attending to features on the red side of the spectrum, and very little attention focused on the region around 5000\,\AA.
A possible interpretation for the abundance of channels attending to the red side, which appears to be very common for the galaxies we inspected, is that most skylines are located there.
As we do not remove them or provide the weights and masks to the encoder, attention channels attuned to them or their residuals may be helpful so that they are \emph{not} mistaken for physically relevant features and propagated to the latent space \citep[cf.][]{Laakom2021}.

Even though this galaxy shows significant H$\alpha$ emission, the channel attending to H$\alpha$ is not readily apparent.
For broader features, e.g. the 4000\,\AA break, the activation pattern would be even less obvious.
To automate the identification of specific attention channels responsible for \emph{any} predicted \spender quantity, we would need to know not just whether the attention channel has been activated, but also whether it is important for the relevant aspect of the model.
To this end, we modify a visualization technique for image classifications called Gradient-weighted Class Activation Mapping \citep[Grad-CAM,][]{Selvaraju2020-hz}. 
In short, we run a forward pass through the autoencoder and retain the attention weights $\ba(\by)$.
Next, we define a scalar function $l(\cdot)$ of the model prediction, e.g. the restframe flux $\bx^\prime$ above continuum $\mathbf{c}$ of the spectral element nearest to the center of the H$\alpha$ line at 6565\,\AA, $l_{\mathrm{H}\alpha}(\bx^\prime)= (\bx^\prime - \mathbf{c})|_{6565}$.
We can then compute the backward gradient $\nabla_{\ba} l(\bx^\prime)$, i.e. the dependency of $l$ on any of the attention weights.
Multiplying these gradients with the attention weights computed in the forward pass,
\begin{equation}
\label{eq:grad-fam}
    \mathbf{g}\ba\equiv\left(\nabla_{\ba}l(\bx^\prime)\right) \odot \left(\ba(\by)\right),
\end{equation}
identifies the activated channel and segment that is most relevant for the value of $l$.

We show this method, which we dub Gradient-weighted Feature Attention Mapping (Grad-FAM), applied to the same galaxy in the top right panel \autoref{fig:attn}.
We can see that channel 149 is most important for the predicted value of H$\alpha$ emission, and that its maximum wavelength segment is that in which the H$\alpha$ line falls.
The bottom left panel of \autoref{fig:attn} shows the same visualization for a higher redshift galaxy with similarly strong H$\alpha$ emission.
The maximum $\mathbf{g}\ba$ is still in channel 149, but it---like the H$\alpha$ line---has moved to a larger wavelength segment, which demonstrates clearly the crucial capability of identifying spectral features regardless of their location in the observed spectrum, i.e. regardless redshift.
Without this feature, finding latent summaries of the restframe spectrum in a redshift-invariant way would not be possible.

We can further exploit the association of H$\alpha$ with attention channel 149.
The bottom right panel of \autoref{fig:attn} shows the restframe models of a 6 randomly selected galaxies from the same SDSS plate, ordered by  $\mathbf{g}\ba_{149}$.
We can clearly see that H$\alpha$ line strength monotonically increases from top to bottom.
Constructing and computing $\mathbf{g}\ba$ tells us how where spectral features are located that are important for a specific aspect of the \spender prediction, and how strong they are.

\subsection{Structure of the Latent Space}
\label{sec:latents}

\begin{figure*}[t]
    \includegraphics[width=\textwidth]{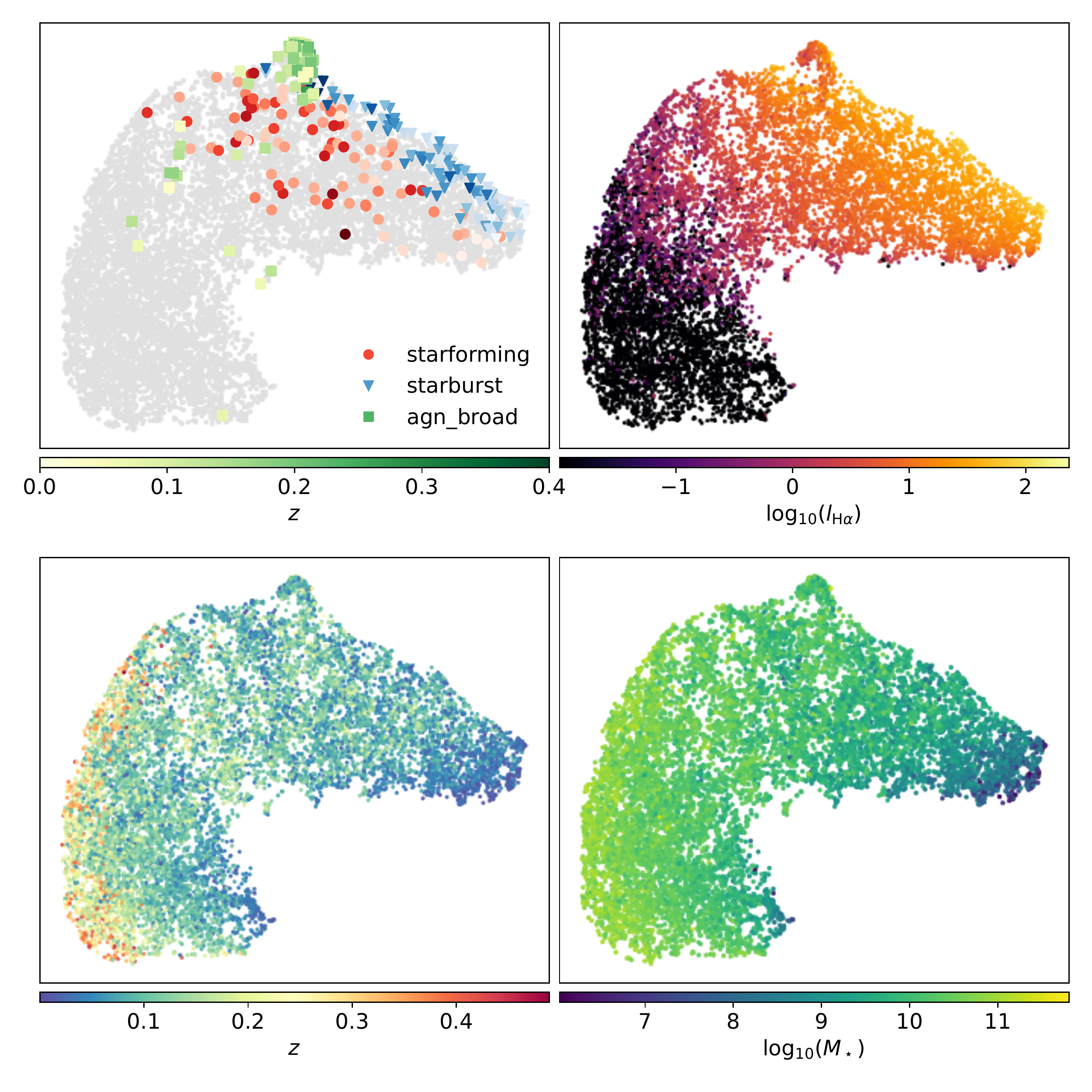}
    \caption{UMAP embedding of the \spender LR-10 latent variables for 20,480 spectra. Galaxy subpopulations (starburst, starforming, and broadline AGN) form clusters in latent space (\emph{top left}, bolder colors indicate higher redshifts). Color coding by H$\alpha$ line strength (\emph{top right}) and stellar mass (\emph{bottom right}) reveals the two main dependencies of the spectrum embeddings, with the redshift dependence (emph{bottom left}) being a result of the magnitude cut of the survey.} 
    \label{fig:umap}
\end{figure*}

To further test our assertion that the \spender architecture produces robust latent representation of an extended restframe model from noisy and redshifted spectra, we embed the 10-dimensional latent space into a 2-dimensional UMAP \citep{McInnes2018-ed}.
\autoref{fig:umap} shows the resulting embedding of 20,480 spectra.
We can see a smooth, compact distribution with two main lobes.
Selecting 100 examples of galaxy types with distinctive spectra---star-forming, starburst, and broadline AGNs%
\footnote{We follow the definition of \texttt{SUBCLASS} in the SDSS DR16 data base, which classified a galaxy as as star-forming if its spectrum has detectable emission lines with $\log({\rm OIII}/{\rm H}\alpha) < 0.7 - 1.2 (\log({\rm NII}/{\rm H}\alpha) + 0.4$. Star-forming galaxies with H$\alpha$ equivalent widths greater than $50$\,\AA\ are classified as starbursts. A galaxy is classified as an AGN if it has detectable emission lines with $\log({\rm OIII}/{\rm H}\alpha) > 0.7 - 1.2(\log({\rm NII}/{\rm H}\alpha) + 0.4)$. Broadened line emission requires velocity dispersion of $\sigma > 200$\,km/s.}%
---shows the expected clustering of similarly typed galaxies.
Within each group, there is no apparent dependence on redshift or SNR, indicating a robust encoding.

The grouping of strong line emitters towards to upper edge suggests that the vertical ordering is related to star formation.
Color-coding each galaxy with the excess over continuum of the H$\alpha$ line $\l_{\mathrm{H}\alpha}$ (see \autoref{sec:attention}) in the top right panel of \autoref{fig:umap} clearly shows the suspected dependence.
A visual inspection further confirms that the lower lobe is entirely composed of quiescent galaxies.
The \spender latents capture the continuum in star-formation rates as one main degrees of freedom in galaxy spectra.
This finding is fully consistent with numerous theoretical and empirical studies, including \citet{Portillo2020-yr} and \citet{Teimoorinia2022} who employed different AE architectures.

This leaves at least one additional degree of freedom.
Color-coding by SNR or reconstruction loss does not show any noticeable dependence of the latents.
However, as the bottom left panel of \autoref{fig:umap} reveals, the horizontal position in the UMAP is clearly correlated with redshift.
While the center of the UMAP is mostly composed of the galaxies from the most common redshift range for the MGS, $0.05 \lesssim z \lesssim 0.25$, and shows no apparent redshift dependence, the left and right edges are populated by the highest and lowest redshift galaxies.
This is an unexpected finding because our architecture was designed to establish a redshift-independent restframe encoding.

We suspect that the data themselves could induce a redshift dependence, but we expect only little actual evolution in the galaxy population over this redshift range.
Selection effects would change the relative abundance of samples in latent space, presumably as a function of redshift, but not their position.
However, in a magnitude-limited sample any dependence on stellar mass $M_\star$ could mimic a redshift dependence because at high redshifts only massive galaxies get targeted, and low-mass galaxies are only targeted at low redshifts.
An increase in halo mass, as traced by stellar mass, leads to a broadening of emission or absorption lines, which the encoder can recognize.
Color-coding the latents by the median of the stellar mass probability distributions for each galaxy from the SDSS MPA-JHU catalog \citep{Kauffmann2003} indeed shows that the horizontal direction of the UMAP is correlated with stellar mass (lower right panel of \autoref{fig:umap}).
This finding can explain the redshift dependence and determine stellar mass as the second major degree of freedom in SDSS-II spectra.
At this point, we cannot rule out a residual redshift dependence, and will investigate this concern in a companion paper (Liang et al., in prep.).

\section{Conclusion and Outlook}

We have introduced the novel neural network architecture \spender as a core building block to represent galaxy spectra with a differentiable, data-driven approach.
Its main novelty is not the use of an autoencoder for this purpose---although the very capable encoding architecture from \citet{Serra2018-lw} has not been used in astronomy before.
The main novelty lies in the separation of the neural restframe representation from the mapping to observational data.
With this architecture, we combine an unsupervised model of the relevant features in galaxy spectra, for which our theoretical modeling abilities are insufficient, with an explicit analytic treatment of the transformations due to varying redshift and instrumental capabilities.
We can directly ingest large quantities of spectroscopic data without the need to de-redshift them or mask artifacts, as the architecture learns to recognize these perturbations during training.

\spender produces highly realistic galaxy spectra over the full range of observed redshifts and noise levels in the SDSS MGS.
The restframe model exceeds the wavelength range and spectral resolution of the original data, and can even be made to deconvolve from the LSF to achieve effective super-resolution.
With a latent space of 6--10 dimensions, it successfully recovers complicated behavior, e.g. of the H$\alpha$-\nii\ complex, for virtually all spectra, even those with substantial line broadening.

The encoder produces an interpretable latent space, whose main ordering is directly related to stellar mass and the amount of line emission in the spectra.
Distinct galaxy types occupy distinct regions in latent space.
Whether the \spender latents are fully redshift invariant as intended from its design will be the subject of a forthcoming study (Liang et al., in prep.).

With such an interpretable low-dimensional latent space, realistic mock spectra can be generated by sampling from or interpolating between the full distribution or regions associated with specific galaxy subtypes.
Transforming the pretrained latent space into a simpler distribution through a normalizing flow model \citep{Papamakarios2021-uz}, which could be conditioned on redshift and other relevant parameters, provides a more principled and computationally efficient approach for generating such mock spectra.
Doing so establishes a fully probabilistic treatment of galaxy spectra, allowing not only sampling but also evaluating the likelihood of given observations.
We intend to exploit the learned distribution of observable spectra to detect outliers and stabilize under-constrained inverse problems arising in super-resolution and data fusion applications.

One main assumption we have made so far is that an accurate redshift estimate is available for every galaxy.
It is evidently possible to estimate the redshift directly from the spectrum, but our encoding architecture is almost uniquely unsuitable for this tasks: the attention module is meant to create representations invariant under translation, which removes the most obvious signature of redshift.
However, \spender implements a differentiable path for the redshift dependence of the loss.
Coupling it with another neural network that performs the redshift estimation is thus a promising avenue for a fully data-driven spectrum analysis pipeline.

To enable reproduction of our findings and aid further development, we make our code and the best-fitting models for SDSS spectra in both low and high-resolution settings publicly available at \url{https://github.com/pmelchior/spender}.

\section*{Acknowledgments}
This work was supported by the AI Accelerator program of the Schmidt Futures Foundation.
The author(s) are pleased to acknowledge that the work reported on in this paper was substantially performed using the Princeton Research Computing resources at Princeton University which is consortium of groups led by the Princeton Institute for Computational Science and Engineering (PICSciE) and Office of Information Technology's Research Computing.

\vspace{1em}
\software{
    \href{https://pytorch.org/}{\code{Pytorch}} \citep{pytorch},
    \href{https://github.com/huggingface/accelerate}{\code{accelerate}} \citep{accelerate},
    \href{https://github.com/aliutkus/torchinterp1d}{\code{torchinterp1d}},
    \href{http://www.numpy.org}{\code{NumPy}} \citep{numpy},
    \href{https://www.astropy.org/}{\code{Astropy}} \citep{astropy},
    \href{https://matplotlib.org}{\code{Matplotlib}} \citep{matplotlib}
}

\bibliography{references}

\end{document}